\documentclass[entropy,review,accept,pdftex,moreauthors]{Definitions/mdpi}

\firstpage{1}
\pubvolume{25}
\issuenum{6}
\articlenumber{944}
\pubyear{2023}
\copyrightyear{2023}
\externaleditor{Academic Editors: Orlando Luongo and Hernando Quevedo}
\datereceived{22 May 2023}
\daterevised{11 June 2023} 
\dateaccepted{13 June 2023}
\datepublished{15 June 2023}
\hreflink{https://doi.org/\linebreak 10.3390/e25060944} 

\Title{Irreversible Geometrothermodynamics of Open Systems in Modified Gravity}

\TitleCitation{Irreversible Geometrothermodynamics of Open Systems in Modified Gravity}

\Author{Miguel A. S.  Pinto
~$^{1,2,\dagger}$\orcidA{}, Tiberiu Harko~$^{3,4,5,\dagger}$\orcidB{} and Francisco S. N. Lobo
~$^{1,2,}$*$^{,\dagger}$\orcidC{}}

\AuthorNames{Miguel A. S.  Pinto, Tiberiu Harko and Francisco S. N. Lobo}

\AuthorCitation{Pinto, M.A.S.; Harko, T.; Lobo, F.S.N.}

\address{%
$^{1}$ \quad Instituto de Astrof\'{\i}sica e Ci\^{e}ncias do Espa\c{c}o, Faculdade de
Ci\^encias da
~Universidade de Lisboa, \mbox{Campo Grande}, Edif\'{\i}cio C8,
P-1749-016 Lisbon, Portugal; mapinto@fc.ul.pt\\
$^{2}$ \quad Departamento de F\'{i}sica, Faculdade de Ci\^{e}ncias da Universidade de Lisboa, Campo Grande, Edif\'{\i}cio C8, P-1749-016 Lisbon, Portugal\\
$^{3}$ \quad Department of Physics, Babes-Bolyai University, Kogalniceanu Street,
	400084 Cluj-Napoca, Romania; tiberiu.harko@aira.astro.ro\\
$^{4}$ \quad Department of Theoretical Physics, National Institute of Physics
and Nuclear Engineering (IFIN-HH), \mbox{077125 Bucharest, Romania}\\
$^{5}$ \quad Astronomical Observatory, 19 Ciresilor Street,
	400487 Cluj-Napoca, Romania}

\corres{Correspondence: fslobo@fc.ul.pt}

\firstnote{These authors contributed equally to this work.}

\abstract{In this work, we explore the formalism of the irreversible thermodynamics of open systems and the possibility of gravitationally generated particle production in modified gravity. More specifically, we consider the scalar--tensor representation of $f(R,T)$ gravity, in which the matter energy--momentum tensor is not conserved due to a nonminimal curvature--matter coupling. In the context of the irreversible thermodynamics of open systems, this non-conservation of the energy--momentum tensor can be interpreted as an irreversible flow of energy from the gravitational sector to the matter sector, which, in general, could result in particle creation. We obtain and discuss the expressions for the particle creation rate, the creation pressure, and the entropy and temperature evolutions. Applied together with the modified field equations of scalar--tensor $f(R,T)$ gravity, the thermodynamics of open systems lead to a generalization of the $\Lambda$CDM cosmological paradigm, in which the particle creation rate and pressure are considered effectively as components of the cosmological fluid energy--momentum tensor.
Thus, generally, modified theories of gravity in which these two quantities do not vanish provide a macroscopic phenomenological description of particle production in the cosmological fluid filling the Universe and also lead to the possibility of cosmological models that start from empty conditions and gradually build up matter and entropy.}

\keyword{geometrothermodynamics; irreversible thermodynamics; modified gravity}

\begin{document}


\section{Introduction}

The study of irreversible matter creation in the context of cosmology started with the pioneering work by Prigogine and his collaborators in the late 1980s \cite{Prigogine:1988,Prigogine:1989zz,Prigogine:1986}. In these works, an alternative cosmology is presented in which the description of large-scale particle and entropy production is based on the reinterpretation of the matter energy--momentum tensor by adding a matter creation term in the adiabatic conservation laws present in the $\Lambda$CDM model. The addition of this matter creation term makes it possible for cosmological matter to be created from an energy flow coming from the gravitational fields.
In the following years, Calvão, Lima, and Waga generalized Prigogine's results by introducing a covariant formulation for the thermodynamic quantities, in particular, the entropy and particle four-flux vectors \cite{Calvao:1991wg}. Currently, the irreversible thermodynamics of open systems is a broadly studied field due to its usefulness in several applications, one of them being the study of irreversible particle creation in cosmologies with an underlying theory of gravity that contains (at least) a nonminimal geometry--matter coupling, namely, in the context of curvature--matter \cite{Harko:2011kv,Harko:2014pqa,Harko:2015pma,Haghani:2013oma,Harko:2014gwa,Harko:2020ibn,Harko:2012ar,Lobo:2022aop}, torsion--matter \cite{Harko:2014aja,Harko:2014sja,Harko:2021bdi}, and non-metricity--matter couplings \cite{Harko:2018gxr}.

However, when using the formalism of thermodynamics of open systems in this context, an obvious question arises: is the Universe an open system? This is a legitimate query since irreversible matter creation processes can only take place in open systems. The Universe is typically considered to be an isolated system because, as far as it is known, it does not receive any kind of energy from outside, for instance, another Universe. Therefore, we must be extremely careful in explaining why it is possible to treat the Universe as an open system. First of all, a general open system consists of a bulk, which is the system's core, and its surroundings, with a boundary tracing the line between the two. At  first sight, one possible way to consider the Universe as an open system would be to assume the Universe itself as the bulk, with its eventual physical limits as the boundary and its surroundings as something outside of it. However, for the reasons stated previously, it would be highly speculative if we do that treatment. Instead, in a co-moving frame of reference, we consider the observable Universe as the bulk, limited by a boundary, which is usually considered to be the apparent horizon (for example, see \cite{Pavon:2012qn,Mimoso:2018juu}) because the thermodynamic laws in accelerated expanding Universes are satisfied in it \cite{Wang:2005pk}, and the unobservable Universe as the surroundings. Thus, the creation of cosmological matter via gravitationally induced particle production processes takes place in the bulk, which exchanges gravitational energy with the surroundings. Hence, the ``Universe'' one refers to when using the formalism of irreversible thermodynamics of open systems is the observable Universe. The last point that needs to be stated is related to the terminology of open systems: generally, one refers to the bulk as the open system to further simplify communication. We shall also adopt this~approach.

In this work, we explore the formalism of irreversible thermodynamics of open systems, and, by using it in the cosmological context, we investigate the possibility of gravitationally generated particle production in modified gravity. More specifically, we consider the equivalent scalar--tensor representation of $f(R,T)$ gravity \cite{Harko:2011kv},  where $R$ is the Ricci scalar and $T$ is the trace of the energy--momentum tensor and in which the latter is not conserved due to a nonminimal curvature--matter coupling. By taking into account the formalism of irreversible thermodynamics of open systems, this non-conservation of the energy--momentum tensor can be interpreted as an irreversible flow of energy from the gravitational sector to the matter sector, which, in general, could result in particle creation. Here, we present and discuss the expressions for the particle creation rate, the creation pressure, and the entropy and temperature evolutions. Applied together with the modified field equations of scalar--tensor $f(R,T)$ gravity, the thermodynamics of open systems lead to a generalization of the standard $\Lambda$CDM cosmological paradigm, in which the particle creation rate and pressure are considered effectively as components of the cosmological fluid energy--momentum tensor \cite{Pinto:2022tlu}.

This paper is organized as follows. In Section~\ref{sec2}, we present the thermodynamics of matter production in the context of open systems and provide a phenomenological description for particle production effects at a macroscopical level. In Section~\ref{sec3}, we apply of the formalism of thermodynamics of irreversible matter creation to cosmology and obtain the entropy and temperature temporal evolutions. In Section~\ref{sec4}, we explore gravitationally induced particle production within the scope of the nonminimal curvature--matter couplings, in particular in $f(R,T)$ gravity, where the covariant divergence of the matter energy--momentum tensor does not vanish. In Section~\ref{sec5}, we present and interpret the phenomenology of the cosmological field equations of the scalar--tensor representation of $f(R,T)$ gravity describing the creation of matter in the formalism of irreversible thermodynamics of open systems. Finally, in Section~\ref{secconcl}, we discuss our results and conclude.

\section{Thermodynamic Interpretation of Matter Creation}\label{sec2}

Having clarified the possible issue of open systems in the Introduction, let us consider a system of this nature with volume $V$ containing $N$ particles, with an energy density $\rho$ and a thermodynamic pressure $p$. For such a system, the first law of thermodynamics, written in its most general form, is given by
\begin{equation}\label{eq:1st law 1}
\mathrm{d}(\rho V)=\mathrm{d} Q-p \mathrm{d} V+\frac{h}{n} \mathrm{d}(n V),
\end{equation}
where $\mathrm{d}Q$ is the heat received by the system during a time $\mathrm{d}t$, $h=\rho + p$ is the enthalpy per unit volume, and $n=N/V$ is the particle number density. Unlike isolated or closed systems, where the number of particles remains constant, the thermodynamic conservation of energy in open systems contains a term that expresses the matter creation/annihilation processes that can occur within the system. Furthermore, the second law of thermodynamics imposes the following constraint on the total entropy $\mathcal{S}$ of any physical system:
\begin{equation}
\label{eq:entropy1}
\mathrm{d}  \mathcal{S}=\mathrm{d}_{e} \mathcal{S}+\mathrm{d}_{i} \mathcal{S}\geq0,
\end{equation}
where $\mathrm{d}_e\mathcal{S}$ is the entropy flow and $\mathrm{d}_i\mathcal{S}$ is the entropy creation. These two terms have different physical meanings as they express two distinct contributions to entropy production: while the first can be seen as the contribution of how the system is arranged, which, in other words, means it measures the change in the system's homogeneity, the latter is the entropy originating from particle creation processes that occur within the open system. To find expressions for these two quantities, we start by writing the total differential of the~entropy:
\begin{equation}
\label{eq:total_diff_entropy1}
\mathcal{T} \mathrm{d} \mathcal{S}=\mathrm{d}(\rho V)+p \mathrm{d} V-\mu \mathrm{d}(n V),
\end{equation}
where $\mathcal{T}$ is the temperature, $\mu$ is the chemical potential, and $s=\mathcal{S}/V$ is the entropy density. These thermodynamic quantities are defined to be positive. By using Equation~\eqref{eq:1st law 1} and the thermodynamic relation
\begin{equation}
    \mu n=h-\mathcal{T}s,
\end{equation}
it is possible to write Equation~\eqref{eq:total_diff_entropy1} in a more useful manner:
\begin{equation}
\mathcal{T} \mathrm{d} \mathcal{S}=\mathrm{d} Q+\mathcal{T} \frac{s}{n} \mathrm{d}(n V).
\end{equation}
Since Equation~\eqref{eq:entropy1} implies that
\begin{equation}
\mathcal{T} \mathrm{d} \mathcal{S}=\mathcal{T} \mathrm{d}_{e} \mathcal{S}+\mathcal{T} \mathrm{d}_{i} \mathcal{S},
\end{equation}
we conclude that the entropy flow and the entropy creation are given by the following expressions, respectively:
\begin{equation}
\label{eq:entropies}
\mathrm{d}_{e} \mathcal{S}=\frac{\mathrm{d} Q}{\mathcal{T}}, \qquad \mathrm{d}_{i} \mathcal{S}=\frac{s}{n} \mathrm{d}(n V).
\end{equation}
From standard cosmology, it is well-known that a homogeneous Universe does not receive energy in the form of heat due to all physical quantities being independent of spatial position---in particular, there are no temperature gradients. In addition, under standard cosmology, the Universe is considered to be an isolated system. As a consequence, the second law of thermodynamics is reduced to
\begin{equation}
    \label{dS isolated}
    \mathrm{d}\mathcal{S} = \frac{\mathrm{d}Q}{\mathcal{T}}=0,
\end{equation}
which indicates that a homogeneous Universe addressed as an isolated system does not change its entropy.

Now, let us see what happens within the scope of open systems. Since entropy flow can be seen as a measure of the variation in the system's homogeneity, if we consider a permanently homogeneous system, there is no change in the homogeneity of the system, which means that its configuration does not change from the thermodynamics perspective. Therefore, we conclude that the entropy flow vanishes in such a system. As before, one could also argue that a homogeneous system does not receive heat, $\mathrm{d}Q=0$, which implies, by the left equality of Equation~\eqref{eq:entropies}, that $\mathrm{d}_{e} \mathcal{S}=0$. Hence, in homogeneous systems, we expect adiabatic processes to occur, and matter creation is the only source of entropy production. In other words, the creation of matter is the only process that can lead to an increase in the entropy of a homogeneous system, as expressed by the following condition:
\begin{equation}
\label{entropy_matter}
\mathrm{d} \mathcal{S}=\mathrm{d}_{i} \mathcal{S}=\frac{s}{n} \mathrm{d}(n V)\geq0.
\end{equation}
In the cosmological context, this leads to the possibility of having an energy flow from the gravitational sources that produce matter, while the inverse process, that of matter producing gravitational sources, is thermodynamically forbidden. This result is very powerful because not only does it break the equivalence between space--time curvature and matter that takes place, for instance, in GR and in $f(R)$ gravity due to the matter energy--momentum tensor being conserved (implying there are no matter creation sources), but it also distinguishes itself from standard cosmology, where $\mathrm{d}\mathcal{S}=0$.

We now apply the formalism of irreversible matter creation of thermodynamics of open systems to cosmology. Let us consider a flat homogeneous and isotropic universe with volume $V$ containing $N$ particles, an energy density $\rho$ and thermodynamic pressure $p$, which is well-described by the flat Friedmann–Lemaître–Robertson–Walker (FLRW) metric (in Cartesian coordinates):
\begin{equation}
\mathrm{d} s^{2}=-\mathrm{d} t^{2}+a^{2}(t)\left(\mathrm{d} x^{2}+\mathrm{d} y^{2}+\mathrm{d} z^{2}\right),
\end{equation}
where $a(t)$ is the scale factor. As we have seen before, homogeneous systems do not receive heat. Therefore, in such a system, Equation~\eqref{eq:1st law 1} becomes
\begin{equation}
\label{conservation_thermo2}
\mathrm{d}(\rho V)+p \mathrm{d} V-\frac{h}{n} \mathrm{d}(n V)=0.
\end{equation}

Moreover, by expressing the volume $V$ in terms of the scale factor $V=a^3(t)$, the thermodynamic conservation equation, Equation~\eqref{conservation_thermo2}, can be written in terms of (total) time derivatives of the physical quantities as
\begin{equation}
\label{eq:1st law 2}
\frac{\mathrm{d}}{\mathrm{d} t}\left(\rho a^{3}\right)+p \frac{\mathrm{d}}{\mathrm{d} t} a^{3}=\frac{\rho+p}{n} \frac{\mathrm{d}}{\mathrm{d} t}\left(n a^{3}\right).
\end{equation}
Applying the time derivatives, one can rewrite \eqref{eq:1st law 2} in an equivalent form
\begin{equation}
\label{eq:1st law 3}
\dot{\rho}+3 H(\rho+p)=\frac{\rho+p}{n}(\dot{n}+3 H n),
\end{equation}
where $H=\dot{a}/a$ is the Hubble function and the overdot denotes the time derivative. Hence, Equation~\eqref{eq:1st law 3} tells us that in this cosmological system, the ``heat'' received is only due to the variation in the particle number density $n$. The time variation of the particle number density in a homogeneous and isotropic geometry is obtained as \cite{Harko:2015pma}
\begin{equation}\label{eq: number density variation}
    \dot{n}+3 H n = \Gamma n,
\end{equation}
with $\Gamma$ being the particle creation rate, defined as $\Gamma \equiv \dot{N}/ N$. Substituting Equation~\eqref{eq: number density variation} into Equation~\eqref{eq:1st law 3}, we obtain the energy conservation equation in an alternative form
\begin{equation}\label{eq:HGamma}
\dot{\rho}+3 H(\rho+p)=(\rho+p)\Gamma.
\end{equation}

For adiabatic transformations describing irreversible particle creation
in an open thermodynamic system, the first law of thermodynamics can be rewritten as an effective energy conservation equation \cite{Prigogine:1988}
\begin{equation}
\frac{\mathrm{d}}{\mathrm{d} t}\left(\rho a^{3}\right)+\left(p+p_{c}\right) \frac{\mathrm{d}}{\mathrm{d} t} a^{3}=0,
\end{equation}
where $p_c$ is the creation pressure, a supplementary pressure considered in open systems due to irreversible matter creation processes. As the name suggests, this quantity provides a phenomenological description for particle production effects (at a macroscopical level). Expressing the equation above in an equivalent manner
\begin{equation}
\label{eq:effective_energy_conservation}
\frac{\mathrm{d}}{\mathrm{d} t}\left(\rho a^{3}\right)+p \frac{\mathrm{d}}{\mathrm{d} t} a^{3}=-p_{c}\frac{\mathrm{d}}{\mathrm{d} t}a^3 ,
\end{equation}
and comparing it with Equation~\eqref{eq:1st law 2}, we can write an expression for the creation pressure
\begin{equation}
    p_c=-\frac{\rho+p}{n}\frac{\mathrm{d}(na^3)/\mathrm{d}t}{\mathrm{d}a^3/\mathrm{d}t},
\end{equation}
which, after some simplifications, takes the following form:
\begin{equation}
\label{eq: creation pressure1}
p_{c}=-\frac{\rho+p}{3 H} \Gamma.
\end{equation}
\textls[-25]{Therefore, to determine the creation pressure, it is sufficient to know the particle creation rate.}

The formalism of irreversible matter creation of thermodynamics of open systems applied in cosmology can describe the creation of matter in a homogeneous and isotropic universe as long as the particle creation rate and, consequently, the creation pressure are not zero. For example, as a result of the conservation of the energy--momentum tensor, both GR and $f(R)$ gravity \cite{Sotiriou:2008rp} are incapable of explaining such particle production because, in these theories, one obtains $ \dot{\rho}+3 H(\rho+p)=0$, which indicates that both the creation rate and creation pressure vanish. Thus, modified theories of gravity in which these two quantities do not vanish can provide a phenomenological description of particle creation in the cosmological fluid filling the Universe.

\section{Entropy and Temperature Evolution}\label{sec3}

Another interesting result concerning the application of the formalism of thermodynamics of irreversible matter creation to cosmology is the possibility of obtaining the temporal evolutions of both entropy and temperature. Therefore, it is possible to have a cosmology in which the Universe gradually accumulates entropy as particles are created at a certain temperature. Furthermore, it is also possible to derive the entropy production rate by introducing the entropy flux four-vector, a covariant generalization of the entropy scalar. The entropic force cosmological models,  which are models that postulate forces of entropic nature to explain the accelerated phases of the Universe, also use this formalism. However, it has been shown that most entropic force cosmological models are indistinguishable from a standard $\Lambda$CDM scenario \cite{GoSa}, which makes them avoidable.

Our objective in the first part of this section is to explore the entropy evolution in a homogeneous and isotropic Universe. To do that, one must recall the secnd law of thermodynamics in the context of open systems, Equation~\eqref{eq:entropy1}. It has been said that the condition of homogeneity imposed implies the vanishing of the entropy flow term, i.e., $\mathrm{d}_e\mathcal{S}=0$. In other words, the only contribution to entropy production is the entropy creation term, so that Equation~\eqref{eq:entropy1} reduces to Equation~\eqref{entropy_matter}. Moreover,  following our previous assumptions that the Universe is both homogeneous and isotropic, by taking the total time derivative of Equation~\eqref{entropy_matter} and using the expression for the entropy creation present in Equation~\eqref{eq:entropies}, the co-moving volume written in terms of the scale factor $V=a^3(t)$, the definition of entropy density ($s=\mathcal{S}/V$), and Equation~\eqref{eq: number density variation}, one can obtain the following expression for the entropy's temporal evolution:
\begin{equation}
\label{eq:entropy_evol}
\frac{\mathrm{d} \mathcal{S}}{\mathrm{d} t}=\mathcal{S} \Gamma \geq 0,
\end{equation}
which has the following solution:
\begin{equation}
\label{S(t)}
\mathcal{S}(t)=\mathcal{S}_{0} \text{exp}\left[\int_{0}^{t} \Gamma\left(t^{\prime}\right) \mathrm{d} t^{\prime}\right],
\end{equation}
where $\mathcal{S}_0=\mathcal{S}(0)$ is the constant initial entropy. Therefore, in a homogeneous and isotropic geometry, in the formalism of irreversible matter creation, what causes the time variation of the entropy is the particle creation rate.

The entropy flux four-vector $S^{\mu}$ was introduced in \cite{Calvao:1991wg} and it is defined as
\begin{equation}
\mathcal{S}^{\mu}=n \sigma u^{\mu} \,,
\end{equation}
where $\sigma=\mathcal{S}/N$ is the entropy per particle (or characteristic entropy). Since $\mathcal{S}^{\mu}$ must obey the second law of thermodynamics, we have the following condition:
\begin{equation}
    \nabla_{\mu} \mathcal{S}^{\mu}\geq0,
\end{equation}
which is the second law of thermodynamics written in a covariant formulation. 
Thus, to obtain the entropy production rate due to matter creation processes, we determine the covariant derivative of the entropy flux four-vector
\begin{equation}
\nabla_{\mu} \mathcal{S}^{\mu} = \left(\nabla_{\mu}n\right) \sigma u^{\mu}+n\left(\nabla_{\mu} \sigma \right) u^{\mu}+n\sigma \nabla_{\mu}u^{\mu},
\end{equation}
which, by using $\nabla_{\mu}u^{\mu}=3H$ and $u^{\mu}\nabla_{\mu}=\mathrm{d}/\mathrm{d}t$, assumes the following form:
\begin{equation}
\label{entropy_prod_rate1}
\nabla_{\mu} \mathcal{S}^{\mu}=(\dot{n}+3 H n)\sigma + n\dot{\sigma}.
\end{equation}

\textls[-25]{To further simplify the expression above, we take the time derivative of the Gibbs~relation~\cite{Calvao:1991wg}}
\begin{equation}
n \mathcal{T} \dot{\sigma}= \dot{\rho}-\frac{\rho + p}{n} \dot{n},
\end{equation}
and use it in combination with the expression for the chemical potential
\begin{equation}
\mu=\frac{h}{n}-\mathcal{T}\frac{s}{n}=\frac{\rho + p}{n}-\mathcal{T} \sigma ,
\end{equation}
alongside Equations~\eqref{eq:1st law 3} and \eqref{eq: number density variation}. With that, we obtain a compact form for the covariant derivative of the entropy flux four-vector
\begin{equation}
\label{entropy_prod_rate}
    \nabla_{\mu} \mathcal{S}^{\mu} = s\Gamma \geq 0.
\end{equation}
It is also possible to explore the similarities between Equations~\eqref{eq:entropy_evol} and \eqref{entropy_prod_rate}. Both the entropy temporal evolution and the entropy production rate depend on the particle creation rate, evidencing the fundamental role played by this quantity in the description of a homogeneous and isotropic universe in which matter creation processes occur. The only difference between the two is that the entropy production rate depends on the entropy density (as expected since we have a flux) while the entropy temporal evolution depends on the entropy itself.

Similarly to entropy, under the formalism of the thermodynamics of open systems, it is possible to obtain an expression for temperature as a function of time for a given theory of gravity. A thermodynamic system is fundamentally described by the particle number density $n$ and the temperature $\mathcal{T}$. Thus, in a thermodynamic equilibrium situation, the energy density $\rho$ and the pressure $p$ are determined from the equilibrium equations of state:
\begin{equation}
\label{drho_and_dp}
\rho=\rho(n, \mathcal{T}), \qquad
p=p(n, \mathcal{T}).
\end{equation}
\textls[-25]{Then, the differential of the energy density and the differential of the pressure are, respectively}:
\begin{equation}
\label{drho}
\mathrm{d} \rho=\left(\frac{\partial \rho}{\partial n}\right)_{\mathcal{T}} \mathrm{d} n+\left(\frac{\partial \rho}{\partial \mathcal{T}}\right)_{n} \mathrm{d} \mathcal{T},
\end{equation}
\begin{equation}
\mathrm{d} p=\left(\frac{\partial p}{\partial n}\right)_{\mathcal{T}} \mathrm{d} n+\left(\frac{\partial p}{\partial \mathcal{T}}\right)_{n} \mathrm{d} \mathcal{T},
\end{equation}
where the subscripts $\mathcal{T}$ and $n$ on the partial derivatives indicate that $\mathcal{T}$ and $n$ are fixed, respectively.
Substituting Equation~\eqref{drho}  in Equation~\eqref{eq:HGamma}, we obtain
\begin{equation}
\label{energy_conserv_temp}
\left(\frac{\partial \rho}{\partial n}\right)_{\mathcal{T}} \dot{n}+\left(\frac{\partial \rho}{\partial \mathcal{T}}\right)_n \dot{\mathcal{T}}+3(\rho+p) H=(\rho+p) \Gamma.
\end{equation}

To express the energy conservation equation above in a more convenient manner, first, we use the Gibbs relation \cite{Calvao:1991wg} to write the differential of the characteristic entropy $\sigma$ as
\begin{equation}
\label{dsigma1}
\mathrm{d}\sigma=\frac{1}{n \mathcal{T}} \mathrm{d} \rho-\frac{\rho+p}{n^{2} \mathcal{T}} \mathrm{d}n.
\end{equation}
By looking at Equation~\eqref{dsigma1}, one could say that $\sigma$ is a function of $\rho$ and $n$. However, since $\rho$ itself is a function of $n$ and $\mathcal{T}$ (Equation~\eqref{drho_and_dp}),  $\sigma$ is thus, fundamentally, a function of $n$ and $\mathcal{T}$. By this reasoning, the true differential of the characteristic entropy is
\begin{equation}
\label{dsigma_2}
\mathrm{d} \sigma=\left(\frac{\partial \sigma}{\partial n}\right)_{\mathcal{T}} \mathrm{d} n+\left(\frac{\partial \sigma}{\partial \mathcal{T}}\right)_{n} \mathrm{d} \mathcal{T}.
\end{equation}
To obtain an explicit expression for this differential, one substitutes Equation~\eqref{drho} into Equation~\eqref{dsigma1}, which yields
\begin{equation}
    \mathrm{d}\sigma=\left[\frac{1}{n\mathcal{T}}\left(\frac{\partial \rho}{\partial n}\right)_{\mathcal{T}}+\frac{\rho + p}{n^2\mathcal{T}}\right] \mathrm{d}n + \frac{1}{n\mathcal{T}}\left(\frac{\partial \rho}{\partial \mathcal{T}}\right)_{n}\mathrm{d}\mathcal{T}.
\end{equation}

The entropy $S$ is an exact differential, and so is the characteristic entropy $\sigma$; therefore, we have the following condition:
\begin{equation}
    \left[\frac{\partial}{\partial \mathcal{T}}\left(\frac{\partial \sigma}{\partial n}\right)_{\mathcal{T}}\right]_{n}=\left[\frac{\partial}{\partial n} \left(\frac{\partial \sigma}{\partial \mathcal{T}}\right)_{n}\right]_{\mathcal{T}}.
\end{equation}
Then, one can obtain the following thermodynamic relation:
\begin{equation}
\left(\frac{\partial \rho}{\partial n}\right)_{\mathcal{T}}=\frac{\rho+p}{n}-\frac{\mathcal{T}}{n}\left(\frac{\partial p}{\partial \mathcal{T}}\right)_{n},
\end{equation}
which is plugged into the energy conservation Equation \eqref{energy_conserv_temp}, and with the help of \linebreak \mbox{Equations~\eqref{eq:1st law 3} and \eqref{eq:HGamma}}, we achieve an expression for the temperature evolution
\begin{equation}
\frac{\dot{\mathcal{T}}}{\mathcal{T}}=c_{s}^{2} \frac{\dot{n}}{n},
\end{equation}
where $c_s=\sqrt{\left(\partial p/ \partial \rho\right)_{n}}$ is the speed of sound. Using Equation~\eqref{eq: number density variation}, we write the temperature evolution in terms of the particle creation rate as  follows:
\begin{equation}
\frac{\dot{\mathcal{T}}}{\mathcal{T}}=c_{s}^{2}(\Gamma-3 H),
\end{equation}
which provides the following solution:
\begin{equation}
\label{temp_evol}
\mathcal{T}(t)=\mathcal{T}_{0} \text{exp}\left\{c_s^2 \int_{0}^{t^{\prime}}\left[ \Gamma\left(t^{\prime}\right)-3 H\left(t^{\prime}\right) \right]\mathrm{d} t^{\prime}\right\},
\end{equation}
where $\mathcal{T}_0=\mathcal{T}(0)$ is the constant initial temperature. 

Due to the presence of the Hubble function $H$ in Equation~\eqref{temp_evol}, we find that the temperature evolution of the newly created particles depends on the expansion of the Universe, which is expected. In conclusion, this formalism allows that modified theories of gravity with a non-zero creation rate lead to the possibility of cosmological models that start from empty conditions and progressively build up matter and entropy.

\section{\boldmath{$f(R,T)$} Gravity}\label{sec4}

In the previous sections, we verified that the formalism of irreversible matter creation of thermodynamics of open systems applied in cosmology can describe the creation of matter in a homogeneous and isotropic universe. Here, we  examine the physical consequences of having an explicit nonminimal curvature--matter coupling in a modified theory of gravity, where the covariant divergence of the matter energy--momentum tensor does not vanish. Thus, one could explore gravitationally induced particle production within the scope of these nonminimal curvature--matter couplings, in particular in $f(R,T)$ gravity \cite{Harko:2015pma}. This modified theory of gravity was formulated in 2011 \cite{Harko:2011kv} and since then has been vastly studied in the usual geometrical representation \cite{Houndjo:2011tu,Alvarenga:2013syu,Shabani:2014xvi,Yousaf:2016lls}. Indeed, this theory might to useful to solve some of the problems left unanswered by GR, such as the entropy production problem precisely because of the non-conservation of the matter energy--momentum tensor. Furthermore, the presentation of an equivalent scalar--tensor theory \cite{Rosa:2021teg} has triggered an intensive investigation in that framework \cite{Rosa:2021tei,Goncalves:2021vci,Goncalves:2022ggq,Rosa:2022osy,Goncalves:2023klv}, which makes it very appealing to study at the moment.

We start this section by presenting the modified field equations of the geometrical representation of $f(R,T)$ gravity and its corresponding conservation equation. Then, we consider its scalar--tensor representation, which will be adopted throughout our work due to two reasons. The first one is because, as said above, this theory in its geometrical representation has received much attention since its formulation (while the scalar--tensor representation has not been as deeply explored). The second reason is that it allows for a dynamic system approach, which helps reduce the order of the metric in the field equations and, as such, may introduce some simplicity.

\subsection{Geometrical Representation}\label{sec4a}

Here, we consider $f(R,T)$ gravity in its geometrical representation. In this representation, the gravitational dynamics are ruled by geometric quantities embedded in Riemannian geometry. In the metric formalism, the one we are considering, it is the metric tensor that mediates the gravitational interaction. As usual, we consider that space-time is a four-dimensional Lorentzian manifold $\mathcal{M}$ equipped with a (Lorentzian) metric $g_{\mu\nu}$, on which one defines a set of coordinates $\{x^\mu\}$. We assume that the Lagrangian has the following~form:
\begin{equation}
\label{lagrangian_frt}
    \mathcal{L} = \frac{1}{2 \kappa^{2}} f(R,T) + \mathcal{L}_{\text{m}}\left({g}{_{\mu\nu}}, \Psi\right),
\end{equation}
where $f(R,T)$ is an arbitrary function of the Ricci scalar, $R$, and of the trace of the matter energy--momentum tensor, $T=g^{\mu\nu} T_{\mu\nu}$. We assume that the matter Lagrangian $\mathcal{L}_{\text{m}}$ is only dependent on the metric tensor components, and on a set of non-gravitational matter fields $\Psi$. Moreover, we consider $G=1$, so that $\kappa^2=8 \pi$. The corresponding action of $f(R,T)$ gravity in its geometrical representation has the expression
\begin{equation}
\label{eq:acaogeo}
S=\frac{1}{2 \kappa^{2}} \int_{\mathcal{M}} \sqrt{-g} f(R, T) \mathrm{d}^4 x+\int_{\mathcal{M}} \sqrt{-g} \mathcal{L}_{\text{m}}\left({g}{_{\mu\nu}}, \Psi\right) \mathrm{d}^4 x.
\end{equation}

By looking at Equation~\eqref{eq:acaogeo}, or equivalently at Equation~\eqref{lagrangian_frt}, we verify that the matter Lagrangian $\mathcal{L}_{\text{m}}$ and the $f(R,T)$ function are explicitly separated. Therefore, in $f(R,T)$ gravity, the possible nonminimal curvature--matter couplings are not originated by terms that depend on $\mathcal{L}_{\text{m}}$. Of course, this contrasts with $f(R,\mathcal{L}_{\text{m}})$ gravity \cite{Harko:2010mv}, in which the nonminimal curvature--matter couplings must explicitly depend on $\mathcal{L}_{\text{m}}$ because it is the only matter-related quantity present in the Lagrangian (and action) of the theory. Then, at first sight, such couplings can only arise in $f(R,T)$ gravity due to cross-terms between $R$, a curvature-related quantity, and $T$, a matter-related quantity. Nonetheless, this theory is particularly subtle. For instance, it has been suggested that a function  $ f(R,T) = f_1(R) + f_2(T)$ could give a complete separation between the gravitational and the matter sectors, i.e., geometry (gravity) and matter being minimally coupled \cite{Fisher:2019ekh}. However, in \cite{Harko:2020ivb}, the authors claim that physical misinterpretations were made, and, therefore, even in that case, it is not possible to have gravity and matter minimally coupled. This forces one to reflect further on the role played by the trace $T$ of the matter energy--momentum tensor in this theory. We argue that $T$ is by itself a nonminimal curvature--matter coupling because it is obtained through the contraction between the metric tensor, which is purely related to the geometry of space-time, and the matter energy--momentum tensor, which describes matter. Nevertheless, this issue will be  clarified further when we discuss gravitationally induced particle production in this theory. For now, let us continue presenting the theory.

The variation in the action \eqref{eq:acaogeo} with respect to the metric tensor, which is the only fundamental field in this representation, leads to (see \cite{Harko:2011kv} for details)
\begin{eqnarray}
\delta S=\frac{1}{2\kappa^2} \int_{\mathcal{M}} \sqrt{-g}  \Big[f_{R}(R,T) R_{\mu \nu}-\frac{1}{2} g_{\mu \nu} f(R, T)+\left(g_{\mu \nu}\square -\nabla_{\mu} \nabla_{\nu}\right) f_{R}(R,T) \nonumber \\
-\kappa^{2} T_{\mu \nu}+f_{T}(R,T)\left(T_{\mu \nu}+\Theta_{\mu \nu}\right)\Big] \delta g^{\mu\nu} \mathrm{d}^4 x,
\label{eq:var_acao_1}
\end{eqnarray}
where $\nabla_\mu$ is the covariant derivative and $\square\equiv\nabla^\alpha\nabla_\alpha$ is the D’Alembert operator. The matter energy--momentum tensor is defined as
\begin{equation}
\label{def energy-mom tensor_2000}
T_{\mu \nu}\equiv -\frac{2}{\sqrt{-g}} \frac{\delta\left(\sqrt{-g} \mathcal{L}_\text{m}\right)}{\delta g^{\mu \nu}}.
\end{equation}
We have denoted the partial derivative of the function $f$ with respect to $R$ and $T$, respectively, as
\begin{equation}
f_R(R,T)\equiv \frac{\partial f(R,T)}{\partial R}, \qquad f_T(R,T)\equiv \frac{\partial f(R,T)}{\partial T}.
\nonumber
\end{equation}

The action principle implies, for an arbitrary variation in the metric tensor $\delta g^{\mu\nu}$, that the quantity inside the brackets in Equation~\eqref{eq:var_acao_1} must be zero. Therefore, we obtain the modified $f(R,T)$ gravity field equations
\vspace{-12pt}
\begingroup\makeatletter\def\f@size{9.5}\check@mathfonts
\def\maketag@@@#1{\hbox{\m@th\fontsize{10}{10}\selectfont\normalfont#1}}

\begin{equation}
\label{eq:frt_campo_geometrical}
f_{R}(R,T) R_{\mu \nu}-\frac{1}{2} g_{\mu \nu} f(R, T)+\left(g_{\mu \nu}\square -\nabla_{\mu} \nabla_{\nu}\right) f_{R}(R,T)=\kappa^{2} T_{\mu \nu}-f_{T}(R,T)\left(T_{\mu \nu}+\Theta_{\mu \nu}\right),
\end{equation}
\endgroup
with the auxiliary tensor $\Theta_{\mu \nu}$ defined as
\begin{equation}
\Theta_{\mu \nu} \equiv g^{\alpha \beta} \frac{\delta T_{\alpha \beta}}{\delta g^{\mu \nu}},
\end{equation}
wherein the assumption $\mathcal{L}_{\text{m}}=\mathcal{L}_{\text{m}}({g}{_{\mu\nu}},\Psi)$ leads to
\begin{equation}
\label{eq:theta_tensor}
\Theta_{\mu \nu}=-2 T_{\mu \nu}+\mathcal{L}_{\text{m}} g_{\mu \nu}-2 g^{\alpha \beta} \frac{\partial^{2} \mathcal{L}_{\text{m}}}{\partial g^{\mu v} \partial g^{\alpha \beta}}.
\end{equation}
As such, it is possible to write the matter energy--momentum tensor as
\begin{equation}
T_{\mu\nu}={g}{_{\mu\nu}} \mathcal{L}_\text{m}-2 \frac{\delta \mathcal{L}_{\text{m}}}{\delta g^{\mu\nu}}.
\end{equation}
Henceforth, we will drop the dependence of the $f(R,T)$ function on $R$ and $T$ to simplify the notation. By taking the divergence of Equation~\eqref{eq:frt_campo_geometrical}, we obtain the conservation equation
\begin{equation}
\label{eq:conservation_equation_georep}
\left(\kappa^{2}-f_{T}\right) \nabla^{\mu} T_{\mu \nu}=\left(T_{\mu \nu}+\Theta_{\mu \nu}\right) \nabla^{\mu} f_{T} +f_{T} \nabla^{\mu} \Theta_{\mu \nu}+f_{R} \nabla^{\mu} R_{\mu \nu}-\frac{1}{2} g_{\mu \nu} \nabla^{\mu}f.
\end{equation}
As we can see, the covariant divergence of the matter energy--momentum tensor does not vanish necessarily. Again, we interpret this result as an exchange of energy and momentum between geometry and matter. Next we consider the scalar--tensor representation of the $f(R,T)$ gravity theory, which will be used until the end of this work.

\subsection{Scalar--Tensor Representation}\label{sec4b}

In modified theories of gravity featuring extra scalar degrees of freedom in comparison to GR, one can deduce a dynamically equivalent scalar--tensor representation. As such, we are going to generalize the transition from the geometrical representation to the scalar--tensor representation.

Let us consider the general case of a modified theory of gravity with $N$ geometrical scalar degrees of freedom, whose action takes the following form
\begin{equation}
\label{eq:acao_geral}
S=\frac{1}{2\kappa^2} \int_{\mathcal{M}} \sqrt{-g} f\left(x_{1},\ldots, x_{N}\right) \mathrm{d}^4 x+\int_{\mathcal{M}} \sqrt{-g} \mathcal{L}_{\text{m}}\left({g}{_{\mu\nu}}, \Psi\right) \mathrm{d}^4 x,
\end{equation}
where $f\left(x_{1}, \ldots, x_{N}\right)$ is a function of those $N$ geometrical scalar degrees of freedom, $x_i$. In order to work in the scalar--tensor representation,
we introduce $N$ auxiliary fields $\phi_i$, where each auxiliary field is associated with one geometrical scalar degree of freedom, such that we can rewrite the action \eqref{eq:acao_geral} in the form
\begin{equation}
\label{eq:acaoauxiliar}
S= \frac{1}{2\kappa^2} \int_{\mathcal{M}} \sqrt{-g}\left[f\left(\phi_{1}, \ldots, \phi_{N}\right)+\sum_{i=1}^{N} \frac{\partial f}{\partial \phi_{i}}(x-\phi)_{i}\right] \mathrm{d}^4 x +\int_{\mathcal{M}} \sqrt{-g} \mathcal{L}_{\text{m}}\left({g}{_{\mu\nu}}, \Psi\right) \mathrm{d}^4 x,
\end{equation}
with $(x-\phi)_{i} \equiv x_{i}-\phi_{i}$. The next step is to define the true scalar fields of the theory, i.e., the ones which will be considered as mediators of the gravitational interaction alongside the metric tensor, in the following way
\begin{equation}
\varphi_{i} \equiv \frac{\partial f}{\partial \phi_{i}}.
\end{equation}

With this consideration, it is now possible to write the action in terms of nonminimal couplings between the scalar fields and its corresponding geometrical degrees of freedom
\begin{equation}
\label{eq:acao_geral_st}
S=\frac{1}{2 \kappa^{2}} \int_{\mathcal{M}} \sqrt{-g}\left[\sum_{i=1}^{N} \varphi_{i} x_{i}-V\left(\varphi_{1}, \ldots, \varphi_{N}\right)\right] \mathrm{d}^4 x+\int_{\mathcal{M}} \sqrt{-g}\mathcal{L}_{\text{m}}\left({g}{_{\mu\nu}}, \Psi\right) \mathrm{d}^4 x,
\end{equation}
where $V\left(\varphi_{1}, \ldots, \varphi_{N}\right)$ is the scalar interaction potential defined as
\begin{equation}
V\left(\varphi_{1}, \ldots, \varphi_{N}\right)\equiv \sum_{i=1}^{N} \varphi_{i} \phi_{i}(\varphi_i)-f\left[\phi_{1}(\varphi_1), \ldots, \phi_{N}(\varphi_N)\right].
\end{equation}
Note that this ``recipe'' is only valid when the action of the theory does not contain a coupling term between the matter Lagrangian and some function. In that case, one must introduce at least one more potential, as it was seen in linear $f(R,\mathcal{L}_{\text{m}})$ gravity \cite{Harko:2010mv,Bertolami:2007gv}.

It is important to reinforce that although the geometrical and scalar--tensor representations of a given theory seem to be radically different from each other, they are equivalent, as both describe the same physics. To prove the equivalence between the two representations, we consider the particular case of $f(R,T)$ gravity, not only because it is the underlying theory of our main work but also due to mathematical simplifications. In the $f(R,T)$ case, we have $N=2$ geometrical scalar degrees of freedom, the Ricci scalar, and the trace of the energy--momentum tensor
\begin{equation}
     x_1\equiv R,\qquad
    x_2\equiv T,
 \end{equation}
and consequently, $N=2$ auxiliary fields, which we define as
\begin{equation}
     \phi_1 \equiv \alpha,\qquad
    \phi_2 \equiv \beta.
 \end{equation}

Thus, we write the $f(R,T)$ gravity action in a particular form of action \eqref{eq:acaoauxiliar}, i.e., in terms of the two auxiliary fields $\alpha$ and $\beta$
\begin{equation}
\label{eq:acaoauxiliar_frt}
S= \frac{1}{2 \kappa^{2}} \int_{\mathcal{M}} \sqrt{-g}\left[f(\alpha, \beta)+f_{\alpha}(R-\alpha)+f_{\beta}(T-\beta)\right] \mathrm{d}^4 x+\int_{\mathcal{M}} \sqrt{-g} \mathcal{L}_{\text{m}}\left({g}{_{\mu\nu}}, \Psi\right) \mathrm{d}^4 x,
\end{equation}
where the subscripts $\alpha$ and $\beta$ denote partial derivatives with respect to these variables. By varying the action with respect to $\alpha$ and $\beta$, we obtain, respectively, the equations of motion for each auxiliary field
\begin{equation}
\label{f1}
f_{\alpha \alpha}(R-\alpha)+f_{\alpha \beta}(T-\beta)=0,
\end{equation}
\begin{equation}
\label{f2}
f_{\beta \alpha}(R-\alpha)+f_{\beta \beta}(T-\beta)=0.
\end{equation}
We can also express Equations~\eqref{f1} and \eqref{f2} in a matrix form $A\textbf{x}=0$ as
\begin{equation}
\begin{pmatrix}
f_{\alpha\alpha} & f_{\alpha\beta} \\
f_{\beta\alpha} & f_{\beta\beta}
\end{pmatrix}
\begin{pmatrix}
R-\alpha\\
T-\beta
\end{pmatrix}=0.
\end{equation}
Matrix equations of this form are known to yield a unique solution if and only if det $A\neq0$, which implies the following
\begin{equation}
    f_{\alpha\alpha}f_{\beta\beta}\neq f_{\alpha\beta}^2.
\end{equation}
In such a case, the unique solution is $R=\alpha$ and $T=\beta$. By inserting these results back into Equation~\eqref{eq:acaoauxiliar_frt}, one can verify that this equation reduces to the form of action \eqref{eq:acaogeo}, proving the equivalence between the two representations, and the scalar--tensor representation is well-defined.

Having proved this equivalence, from now on, we will work with the particular case of the general action \eqref{eq:acao_geral_st} for $f(R,T)$ gravity. We define the two dynamical scalar fields as
\begin{equation}
\label{eq:varphi&psi}
     \varphi_1 \equiv \varphi\equiv\frac{\partial f}{\partial R} ,\qquad
    \varphi_2 \equiv \psi\equiv\frac{\partial f}{\partial T},
 \end{equation}
and hence we write the final expression for the $f(R,T)$ action in the scalar--tensor representation as
\begin{equation}\label{eq:STaction}
    S = \frac{1}{2\kappa^2} \int_{\mathcal{M}} \sqrt{-g} \left[\varphi R+\psi T - V(\varphi, \psi)\right]\mathrm{d}^4 x \\
    + \int_{\mathcal{M}} \sqrt{-g} \mathcal{L}_{\text{m}}\left({g}{_{\mu\nu}}, \Psi\right) \mathrm{d}^4 x,
\end{equation}
with the scalar interaction potential defined as
\begin{equation}
\label{eq:interaction potential}
V(\varphi, \psi) \equiv \varphi \alpha(\varphi)+\psi \beta(\psi)-f\left[(\alpha(\varphi), \beta(\psi))\right]
\end{equation}
Similarly to what happens in the equivalent scalar--tensor representation of $f(R)$ gravity, the scalar field $\varphi$ is analogous to a Brans--Dicke scalar field with parameter $\omega_{BD}= 0$. In addition to this scalar field $\varphi$,  the  second  scalar  degree  of  freedom  of $f(R,T)$ gravity, associated with the arbitrary dependence of the action in $T$, is also represented by a scalar field, $\psi$, and together with $\varphi$ have an interaction potential $V(\varphi,\psi)$.

The  action  \eqref{eq:STaction}  depends on three fundamental fields, the metric ${g}{_{\mu\nu}}$ and the two scalar fields $\varphi$ and $\psi$. Varying this action with respect to the metric ${g}{_{\mu\nu}}$ yields the field equation
\begin{equation}
\label{eq:str_field_equation}
\varphi R_{\mu \nu}-\frac{1}{2} g_{\mu \nu}(\varphi R+\psi T-V)+\left(g_{\mu \nu} \square-\nabla_{\mu} \nabla_{\nu}\right) \varphi=\kappa^{2} T_{\mu \nu}-\psi\left(T_{\mu \nu}+\Theta_{\mu \nu}\right).
\end{equation}
Note that this equation is in all respects the same as \eqref{eq:frt_campo_geometrical} but with the partial derivatives of the function $f$ now expressed as scalar fields, as shown in Equation~\eqref{eq:varphi&psi}, and using the definition \eqref{eq:interaction potential} with $\alpha=R$ and $\beta=T$.
Additionally, by taking the variation in Equation~\eqref{eq:STaction} with respect to the scalar fields $\varphi$ and $\psi$, we obtain the equation of motion for $\varphi$ and $\psi$,~respectively
\begin{equation}
V_{\varphi}=R,  \qquad  V_{\psi}=T,
\end{equation}
where the subscripts in $V_{\varphi}$ and $V_{\psi}$ denote the derivatives of the scalar interaction potential $V(\varphi,\psi)$ with respect to the fields $\varphi$ and $\psi$, respectively. By taking the covariant divergence of Equation~\eqref{eq:str_field_equation}, we find the conservation equation for $f(R,T)$ gravity in the scalar--tensor~representation
\begin{equation}
\label{eq:str_divergence}
\left(\kappa^{2}-\psi\right) \nabla^{\mu} T_{\mu \nu}=\left(T_{\mu \nu}+\Theta_{\mu \nu}\right) \nabla^{\mu} \psi+\psi \nabla^{\mu} \Theta_{\mu \nu}-\frac{1}{2} g_{\mu \nu}\left[R \nabla^{\mu} \varphi+\nabla^{\mu}(\psi T-V)\right].
\end{equation}
This equation could also be obtained directly from Equation~\eqref{eq:conservation_equation_georep} by using the geometrical result $\nabla^{\mu}\left(R_{\mu \nu}-\frac{1}{2} g_{\mu \nu} R\right)=0$ and the definitions \eqref{eq:varphi&psi} and \eqref{eq:interaction potential} with $\alpha=R$ and $\beta=T$.

\section{Cosmological Equations}\label{sec5}

The Universe at large scales seems to be homogeneous and isotropic, i.e., following the cosmological principle, and spatially flat. Thus, we consider a Universe described by the flat FLRW metric, which models a Universe with such properties. In the usual spherical coordinates $(t,r,\theta,\phi)$, it takes the form
\begin{equation}
\mathrm{d} s^{2}=-\mathrm{d} t^{2}+a^{2}(t)\left[\mathrm{d} r^{2} +r^{2}\left(\mathrm{d} \theta^{2}+\sin ^{2} \theta \mathrm{d} \phi^{2}\right)\right],
\end{equation}
where $a(t)$ is the scale factor. We also assume that matter is described by a perfect fluid and thus its energy--momentum tensor is given by
\begin{equation}
T_{\mu \nu}=(\rho+p) u_{\mu} u_{\nu}+p g_{\mu \nu},
\end{equation}
where $\rho$ is the energy density, $p$ is the pressure, and $u^{\mu}$ is the four-velocity, which satisfies the normalization condition $u_{\mu}u^{\mu}=-1$. Assuming the matter Lagrangian to be \mbox{$\mathcal{L}_{\text{m}}=p$~\cite{Bertolami:2007gv,Bertolami:2008ab}} and using Equation~\eqref{eq:theta_tensor}, we find an explicit expression for the tensor $\Theta_{\mu\nu}$
\begin{equation}
\Theta_{\mu \nu}=-2 T_{\mu \nu}+p g_{\mu \nu} \,.
\end{equation}

Due to homogeneity, all physical quantities must only depend on the time coordinate $t$, i.e., $\rho=\rho(t)$, $p=p(t)$, $\varphi=\varphi(t)$, and $\psi=\psi(t)$. With these assumptions taken into account, one obtains two independent field equations from Equation~\eqref{eq:str_field_equation}, namely, the modified Friedmann equation and the modified  Raychaudhuri  equation,  which take the following~forms:
\begin{equation}\label{eq:tt}
\dot{\varphi}\left(\frac{\dot{a}}{a}\right)+\varphi\left(\frac{\dot{a}}{a}\right)^2=\frac{8 \pi}{3} \rho+\frac{\psi}{2}\left(\rho-\frac{1}{3} p\right)+\frac{1}{6} V,
\end{equation}
\begin{equation}\label{eq:rr}
\ddot{\varphi}+2 \dot{\varphi}\left(\frac{\dot{a}}{a}\right)+\varphi \left(\frac{2 \ddot{a}}{a}+\frac{\dot{a}^2}{a^2}\right)=-8 \pi p+\frac{\psi}{2}(\rho-3 p)+\frac{1}{2} V,
\end{equation}
respectively, where the overdots denote time derivatives. Combining these two equations, one can obtain new expressions for the equations of motion for the scalar fields $\varphi$ and $\psi$
\begin{equation}
\label{eq:eq_motion_vphi}
V_{\varphi}=R=6\left(\frac{\ddot{a}}{a}+\frac{\dot{a}^2}{a^2}\right),
\end{equation}
\begin{equation}
\label{eq:eq_motion_vpsi}
V_{\psi}=T=3 p-\rho.
\end{equation}
Under the previous assumptions, we determine the cosmological energy conservation equation by fixing $\nu=0$ in Equation~\eqref{eq:str_divergence}, which yields
\begin{equation}\label{eq:conserv-total}
\begin{aligned}
 \dot{\rho}  +   3(\rho+p)\left(\frac{\dot{a}}{a}\right)= \frac{3}{8\pi} \Bigg\{ \dot{\varphi}\left(\frac{\ddot{a}}{a}+\frac{\dot{a}^2}{a^2}-\frac{1}{6}V_\varphi\right)-\dot{\psi}\left(\frac{1}{2}\rho - \frac{1}{6}p+\frac{1}{6}V_\psi\right) \\
     -\psi\left[\frac{\dot{a}}{a}(\rho+p)+\frac{1}{2}\dot{\rho} - \frac{1}{6}\dot{p}\right]  \Bigg\}.
\end{aligned}
\end{equation}

The system of Equations~\eqref{eq:tt}--\eqref{eq:conserv-total} forms a system of five equations from which only four are linearly independent. To prove this feature, one can take the time derivative of Equation~\eqref{eq:tt}; use Equations~\eqref{eq:eq_motion_vphi} and \eqref{eq:eq_motion_vpsi} to eliminate the partial derivatives $V_\varphi$ and $V_\psi$; use the conservation equation in Equation~\eqref{eq:conserv-total} to eliminate the time derivative $\dot\rho$; and use the Raychaudhuri equation in Equation~\eqref{eq:rr} to eliminate the second time derivative $\ddot a$, thus recovering the original equation. Thus, one of these equations can be discarded from the system without loss of generality. Given the complicated nature of Equation~\eqref{eq:rr}, we chose to discard this equation and consider only Equations~\eqref{eq:tt} and \eqref{eq:eq_motion_vphi}--\eqref{eq:conserv-total}.

By including the Hubble function $H=\dot{a}/a$, the set of cosmological equations of scalar--tensor $f(R,T)$ gravity assume the following form
\begin{equation}
    \label{Fr1}
    3H^2=8\pi \frac{\rho}{\varphi}+\frac{3\psi}{2\varphi}\left(\rho-\frac{1}{3}p\right)+\frac{1}{2}\frac{V}{\varphi}-3H\frac{\dot{\varphi}}{\varphi},
\end{equation}
\begin{equation}
    \label{Fr2}
    2\dot{H}+3H^2=-8\pi \frac{p}{\varphi}+\frac{\psi}{2\varphi}\left(\rho-3p\right)+\frac{1}{2}\frac{V}{\varphi}-\frac{\ddot{\varphi}}{\varphi}-2H\frac{\dot{\varphi}}{\varphi},
\end{equation}
\begin{equation}
    \label{Fr3}
    V_{\varphi}=6\left(\dot{H}+2H^2\right), \qquad V_{\psi}=3p-\rho,
\end{equation}
\begin{equation}
    \label{Fr4}
    \dot{\rho}+3H(\rho +p)=\frac{3}{8\pi}\Bigg\{-\frac{\dot{\psi}}{2}\left(\rho-\frac{p}{3}+\frac{V_{\psi}}{3}\right)-\psi\left[H(\rho+p)+\frac{1}{2}\left(\dot{\rho}-\frac{1}{3}\dot{p}\right)\right]\Bigg\}.
\end{equation}

It should be noted that, unlike GR or any modified theory of gravity in which the matter energy--momentum tensor is conserved, the scalar--tensor $f(R,T)$ gravity admits a conservation equation in which the right-hand side of Equation~\eqref{Fr4} does not vanish identically. By recalling Equation~\eqref{eq:HGamma}, such a result implies a particle creation rate $\Gamma\neq0$ and hence the presence of particle production. Substituting $V_{\varphi}$ and $V_{\psi}$ by their expressions, Equations~\eqref{eq:eq_motion_vphi} and \eqref{eq:eq_motion_vpsi}, respectively, we obtain the explicit expression for the particle creation rate in scalar--tensor $f(R,T)$ gravity
\begin{equation}\label{eq:gamma 2}
\Gamma=-\frac{\psi}{8 \pi+\psi}\left(\frac{\mathrm{d}}{\mathrm{d} t} \ln \psi+\frac{1}{2} \frac{\dot{\rho}-\dot{p}}{\rho+p}\right).
\end{equation}
Additionally, by combining Equation~\eqref{eq: creation pressure1} with Equation~\eqref{eq:gamma 2}, we find that the creation pressure in this theory assumes the following expression:
\begin{equation}\label{eq: creation pressure 2}
p_{c}=\frac{\rho+p}{3 H} \frac{\psi}{8 \pi+\psi}\left(\frac{\mathrm{d}}{\mathrm{d} t} \ln \psi+\frac{1}{2} \frac{\dot{\rho}-\dot{p}}{\rho+p}\right).
\end{equation}

As mentioned above, the creation pressure $p_c$ is the (effective) quantity phenomenologically describing the creation of matter in the formalism of irreversible thermodynamics of open systems. As can be seen from Equation~\eqref{eq: creation pressure 2}, $p_c$ has a dependence on one of the two fundamental scalar fields of scalar--tensor $f(R,T)$ gravity, namely, $\psi$. Since this field mediates the gravitational interaction (alongside $g_{\mu\nu}$ and $\varphi$) but also plays an active role in the particle production through $p_c$, we conclude that this production is of gravitational nature. In addition, one should also question why the creation rate and pressure depend only on one of the two scalars. In action \eqref{eq:STaction}, we see that the scalar field $\psi$ is nonminimally coupled with the trace of the energy--momentum tensor $T$. Given that $\psi$ is a gravitational field and $T$ is associated with matter, then we have a direct interaction between gravity and matter. This interaction allows the exchange of energy and momentum between the two sectors, with $\psi$ being the source of matter creation. On the contrary, the scalar $\varphi$ is irrelevant in the process of particle production because it is not nonminimally coupled with a matter-related quantity, only with the Ricci scalar $R$, which is just a geometrical quantity. In conclusion, the scalar $\psi$ is the degree of freedom that triggers particle production in scalar--tensor $f(R,T)$ gravity. Moreover, both Equations~\eqref{eq:gamma 2} and \eqref{eq: creation pressure 2} can be taken as correct because they are consistent with \cite{Harko:2014pqa}, in which these two quantities were obtained in the usual geometrical representation. Consequently, our results prove once again the equivalence between these two representations of $f(R,T)$ gravity.

With the particle creation rate determined, we can now use the general results found in the previous sections regarding the entropy evolution and the temperature evolution to compute these expressions in scalar--tensor $f(R,T)$ gravity. Thus, by substituting \mbox{Equation~\eqref{eq:gamma 2}} in Equation~\eqref{S(t)}, we obtain the explicit expression for the co-moving entropy~evolution
\begin{equation}
\mathcal{S}(t)=\mathcal{S}_{0} \text{exp}\left[-\int_{0}^{t} \frac{\psi}{8 \pi+\psi}\left(\frac{\mathrm{d}}{\mathrm{d} t^{\prime}} \ln \psi+\frac{1}{2} \frac{\dot{\rho}-\dot{p}}{\rho+p}\right) \mathrm{d} t^{\prime}\right],
\end{equation}
while, by substituting Equation~\eqref{eq:gamma 2} in Equation~\eqref{entropy_prod_rate}, we obtain the entropy production rate
\begin{equation}
\nabla_{\mu} \mathcal{S}^{\mu}=-\frac{\psi}{8 \pi+\psi}\left(\frac{\mathrm{d}}{\mathrm{d} t} \ln \psi+\frac{1}{2} \frac{\dot{\rho}-\dot{p}}{\rho+p}\right)s \geq 0.
\end{equation}
In addition, by using Equation~\eqref{eq:gamma 2} together with Equation~\eqref{temp_evol}, we find that the temperature time evolution is given by
\begin{equation}
T(t)=T_{0} \text{exp}\left\{c_s^2 \int_{0}^{t} \left[\frac{\psi}{8 \pi+\psi}\left(\frac{\mathrm{d}}{\mathrm{d} t^{\prime}} \ln \psi+\frac{1}{2} \frac{\dot{\rho}-\dot{p}}{\rho+p}\right) -3H\right] \mathrm{d} t^{\prime}\right\}.
\end{equation}

Furthermore, in order to gain an understanding of how much the Universe decelerates/accelerates throughout its (cosmological) evolution, an indicator is needed. In this sense, it is convenient to introduce the deceleration parameter $q$, which is defined as
\begin{equation}
    q=\frac{\mathrm{d}}{\mathrm{d}t}\frac{1}{H}-1=-\frac{\dot{H}}{H^2}-1.
\end{equation}
Using the cosmological field Equations (\ref{Fr1}) and (\ref{Fr2}), we obtain the expression for the deceleration parameter $q$ in scalar--tensor $f(R,T)$ gravity
\begin{equation}
    q=\frac{1}{2}+\frac{3\left[ 4\pi \frac{p}{\varphi }-\frac{\psi }{4\varphi }%
\left( \rho -3p\right) -\frac{V}{4\varphi }+\frac{\ddot{\varphi}}{2\varphi }%
+H\frac{\dot{\varphi}}{\varphi }\right] }{8\pi \frac{\rho }{\varphi }+\frac{%
\psi }{2\varphi }\left( \rho -3p\right) +\frac{V}{2\varphi }-3H\frac{\dot{%
\varphi}}{\varphi }}.
\end{equation}

\section{Conclusions}\label{secconcl}

In this work, we have explored the formalism of irreversible thermodynamics of open systems and provided a brief overview of the use of this formalism in the cosmological context. By treating a homogeneous and isotropic Universe as an open system, we have seen that the only contribution to entropy production is due to entropy creation, in turn related to particle creation/annihilation processes that occur within the Universe. In addition, the thermodynamic conservation equation of an open system contains an extra term that takes into account the change in the number of particles in the system. Such a term allows us to write this equation in terms of a creation rate but also in terms of a creation pressure within this formalism, with the first being defined as the rate between the time derivative of the number of particles and the number of particles and the second describing the emergence of gravitationally induced macroscopical matter created in space-time. 
 In this respect, we have seen that it is possible to obtain an explicit expression for the creation pressure if one has an explicit expression for the particle creation rate.

Furthermore, we have derived general expressions for the entropy and temperature evolutions by resorting to the second law of thermodynamics and by considering the number density and the temperature as the fundamental thermodynamic quantities that fully describe the open thermodynamic system. We have seen that both evolutions depend on the particle creation rate, which highlights the impact that the creation of matter has on entropy production, with the temperature evolution also depending on the expansion of the Universe according to the presence of the Hubble function. Therefore, the use of the irreversible thermodynamics of open systems in the cosmological domain could help us explain the rise in entropy that occurs during the creation of matter, as long as we consider a modified theory of gravity in which the matter energy--momentum tensor is not conserved, i.e.,  a theory that possesses interaction term(s) between the gravitational fields and~matter.

In this context, considering theories that contain an interaction term between the gravitational fields and the non-gravitational matter fields, either using an explicit nonminimal curvature--matter coupling in the geometrical representation or  a nonminimal coupling between a fundamental scalar field and matter in the scalar--tensor representation, we verify that the matter energy--momentum tensor is not conserved. If we interpret this result as the fact that the fundamental gravitational fields are sources of matter creation, then it is legitimate to resort to the formalism of irreversible matter creation and treat the Universe as an open system. By doing so, we verified that both the particle creation rate and the creation pressure do not vanish identically in the case of scalar--tensor $f(R,T)$ gravity.  Indeed, it seems that there is an evident and profound connection between the non-conservation of the energy--momentum tensor of matter and these two quantities.

Apart from the matter density, thermodynamic pressure, and the Hubble function, we have seen that both the creation rate and the creation pressure only depend on one of the two fundamental scalar fields of the theory, the scalar $\psi$, defined as $\psi \equiv \partial f/ \partial T$. Since this scalar field is one of the three fields that mediate the gravitational interaction in scalar--tensor $f(R,T)$ gravity, we conclude that particle creation is being gravitationally induced. Moreover, it is also possible to analyze this result in the more ``natural'' geometrical representation by using the definition of $\psi$. Hence, it is the partial derivative of $f(R,T)$ with respect to the trace of the matter energy--momentum tensor $T$ that plays the role that the scalar field $\psi$ played in the scalar--tensor representation. Thus, both creation rate and creation pressure are zero if the $f(R,T)$ function does not depend on $T$, and as a consequence, the Ricci scalar $R$ does not contribute to particle creation.

Thus, it is sufficient to have a function that depends only on $T$ due to the fact that this quantity constitutes itself a nonminimal curvature--matter coupling, as it results from a coupling between the metric tensor $g_{\mu\nu}$, which is the only fundamental gravitational field of the theory in the geometrical representation, and the matter energy--momentum tensor $T_{\mu\nu}$. Therefore, it is this degree of freedom by itself that is responsible for the non-conservation of $T_{\mu\nu}$. Then, we can point out that in the geometrical representation of $f(R,T)$ gravity, it is the metric $g_{\mu\nu}$, which participates in the nonminimal curvature--matter coupling encapsulated by $T$, that induces particle production gravitationally.  To conclude, we point out that an extensive discussion regarding the physical nature of the particles that could be created via gravitationally induced creation processes has been carried out in \cite{Harko:2015pma}.

\vspace{6pt}



\newpage
\authorcontributions{Formal analysis, M.A.S.P., T.H. and F.S.N.L.; Investigation, M.A.S.P., T.H. and F.S.N.L.; Writing---original draft, M.A.S.P., T.H. and F.S.N.L. All the authors have substantially contributed to the present work. All authors have read and agreed to the published version of the~manuscript.}

\funding{This research was funded by the Funda\c{c}\~{a}o para a Ci\^{e}ncia e a Tecnologia (FCT) from the research grants UIDB/04434/2020, UIDP/04434/2020, CERN/FIS-PAR/0037/2019, and PTDC/FIS-AST/0054/2021. }

\institutionalreview{Not applicable.}

\informedconsent{Not applicable.}

\dataavailability{Not applicable.}

\acknowledgments{M.A.S.P. acknowledges support from the Funda\c{c}\~{a}o para a Ci\^{e}ncia e a Tecnologia (FCT)  through the Fellowship UI/BD/154479/2022.
The work of T.H. is supported by a grant of the Romanian Ministry of Education and Research, CNCS-UEFISCDI, project number PN-III-P4-ID-PCE-2020-2255 (PNCDI III). 
F.S.N.L. acknowledges support from the Funda\c{c}\~{a}o para a Ci\^{e}ncia e a Tecnologia (FCT) Scientific Employment Stimulus contract with reference CEECINST/00032/2018 and funding from the research grant CERN/FIS-PAR/0037/2019.
M.A.S.P. and F.S.N.L. also acknowledge support from the Funda\c{c}\~{a}o para a Ci\^{e}ncia e a Tecnologia (FCT)  research grants UIDB/04434/2020 and UIDP/04434/2020, and through the FCT project with reference PTDC/FIS-AST/0054/2021 (``BEYond LAmbda'').
}

\conflictsofinterest{The authors declare no conflict of interest.}

\begin{adjustwidth}{-\extralength}{0cm}

\reftitle{References}

\PublishersNote{}
\end{adjustwidth}
\end{document}